\documentstyle[12pt,epsf,eqsection,indent,subeqnarray]{article}

\def\ffrac#1#2{\textstyle{#1\over#2}\displaystyle}

\def\spose#1{\hbox to 0pt{#1\hss}}
\def\ltapprox{\mathrel{\spose{\lower 3pt\hbox{$\mathchar"218$}}
 \raise 2.0pt\hbox{$\mathchar"13C$}}}
\def\gtapprox{\mathrel{\spose{\lower 3pt\hbox{$\mathchar"218$}}
 \raise 2.0pt\hbox{$\mathchar"13E$}}}

\begin{document}
\topmargin 0pt
\oddsidemargin 5mm
\renewcommand{\thefootnote}{\arabic{footnote}}
\newpage
\setcounter{page}{0}
\begin{titlepage}
\vspace{0.5cm}

\begin{center}
{\large \bf Unusual Corrections to Scaling} \\
{\large \bf in the 3-state Potts Antiferromagnet} \\
{\large \bf on a Square Lattice}\\
\vspace{1.8cm}
{\large John Cardy$^{a,d}$, Jesper Lykke Jacobsen$^b$ and
        Alan D.~Sokal$^{c,d}$} \\ 
\vspace{0.5cm}
{\em $^a$ Theoretical Physics, University of Oxford,}\\
{\em 1 Keble Road, Oxford OX1 3NP, United Kingdom}\\
{\em $^b$ Laboratoire de Physique Th\'eorique et Mod\`eles Statistiques,}\\
{\em B\^atiment 100, Universit\'e Paris-Sud, F-91405 Orsay, France}\\
{\em $^c$ Department of Physics, New York University,}\\
{\em 4 Washington Place, New York, NY 10003 USA}\\
{\em $^d$ All Souls College, Oxford} \\
\vspace{0.5cm}
January 11, 2001 \\ slight revisions May 3, 2001
\end{center}
\vspace{1.2cm}

\renewcommand{\thefootnote}{\arabic{footnote}}
\setcounter{footnote}{0}

\begin{abstract}
\noindent
At zero temperature, the 3-state antiferromagnetic Potts model on a
square lattice maps exactly onto a point of the 6-vertex model
whose long-distance behavior is equivalent to that of a free scalar boson.
We point out that at nonzero temperature there are two distinct types
of excitation: vortices, which are relevant with renormalization-group
eigenvalue $\frac12$; and non-vortex unsatisfied bonds,
which are strictly marginal and serve only to renormalize the
stiffness coefficient of the underlying free boson.
Together these excitations lead to an unusual form for the
corrections to scaling: for example, the correlation length diverges as
$\beta\equiv J/kT\to\infty$ according to
$\xi\sim Ae^{2\beta}(1+b\beta e^{-\beta}+\cdots)$,
where $b$ is a nonuniversal constant that may nevertheless be determined
independently. A similar result holds for the staggered susceptibility.
These results are shown to be consistent with the anomalous behavior
found in the Monte Carlo simulations of Ferreira and Sokal.
\end{abstract}

\vspace{.3cm}

\end{titlepage}

\section{Introduction}
\label{sec1}

Corrections to scaling at continuous phase transitions
have long been understood within the framework of the renormalization
group~\cite{Wegner}.
Irrelevant operators give rise to power-law corrections to scaling in,
for example, the temperature-dependence of the correlation length, of the form
\begin{equation}
\xi \;\sim\; A|t|^{-\nu} (1 \,+\, C|t|^\omega \,+\, \ldots)
\end{equation}
where $\omega > 0$ is a universal correction-to-scaling exponent,
and $A$ and $C$ are nonuniversal amplitudes.
(For a critical point at temperature $T_{\rm c} \neq 0$,
the scaling variable $t$ is conventionally defined by $t\propto T-T_{\rm c}$.)
In the case of a marginally irrelevant operator,
both multiplicative and additive logarithmic corrections can occur,
typically of the form
\begin{equation}
\xi \;\sim\; A|t|^{-\nu}(\log |t|^{-1})^{\bar\nu}
   \left[ 1 \,+\, C {\log\log |t|^{-1} \over \log |t|^{-1}} \,+\,
                  C' {1 \over \log |t|^{-1}} \,+\, \ldots \right]
   \;.
\end{equation}

However, it is possible for more exotic dependences to arise. This
paper considers just such an example, namely the antiferromagnetic 3-state
Potts model on a square lattice. This model has the Hamiltonian
\begin{equation}
\label{ham}
H \;=\; J\sum_{\langle ij \rangle}\delta_{s_i,s_j}
\end{equation}
where the sum is over nearest-neighbor pairs of vertices on the
square lattice, at each of which is a degree of freedom $s_i$ taking one
of three possible states. The coupling is antiferromagnetic, that is, $J>0$. 

One reason that this model is exotic is that its critical point occurs
at $T=0$
\cite{Lenard_67,Baxter_70,Baxter_82b,Nijs_82,Kolafa_84,Park_89,%
Burton_Henley_97,Salas-Sokal_98}.
Nevertheless, this is a bona fide critical point,
exhibiting, for example, power-law decay of the order-parameter
correlation functions. At nonzero temperature, these correlation functions
decay exponentially with a finite correlation length $\xi$.
In the past there has been some confusion concerning the correct choice
of scaling variable at such zero-temperature critical points;
it does not necessarily correspond to setting $T_{\rm c}$ to zero
in the above definition of $t$ (i.e.\ to taking $t \propto T$).
In fact, this question is answered quite explicitly,
at least for non-quantum critical points,
by the renormalization group:
the space of scaling variables is the tangent space at the fixed point
to the manifold parametrized by the Boltzmann weights of the model in question. 
In our case, the low-temperature configurations correspond to modifying
one of the ground states by allowing a nonzero density of
nearest-neighbor bonds where $s_i = s_j$
(we henceforth call these {\em unsatisfied bonds}\/).
Each unsatisfied bond occurs with Boltzmann weight $e^{-\beta}$
(where $\beta\equiv J/kT$), and
therefore the correct scaling variable $t$ should be linear in $e^{-\beta}$.

The other, more interesting, reason that this model is unusual is that,
as we shall discuss in detail in the next section, switching on the
temperature excites not just one, but two, scaling operators.
One of these operators turns out to be relevant,
with renormalization-group eigenvalue $y=\frac12$
(hence scaling dimension $x=2-y=\frac32$), corresponding to $\nu=1/y=2$,
so that the leading behavior of the correlation length
is $\xi\propto (e^{-\beta})^{-\nu}=e^{2\beta}$.
The other operator is marginal --- not marginally irrelevant,
but {\em strictly}\/ marginal in the sense that, taken alone,
it would generate a line of fixed points with continuously varying exponents.
The main result of this marginal operator is to give the exponent $\nu$
in the foregoing expression an effective dependence on the scaling variable
$w \equiv e^{-\beta}$ itself: thus we find that
\begin{equation}
\label{corr}
\xi\sim A(e^{-\beta})^{-\nu_{\rm eff}(e^{-\beta})}
\sim Ae^{2\beta}(1+b\beta e^{-\beta} +\cdots)
\end{equation}
where $b = d\nu_{\rm eff}(w)/dw|_{w=0}$ is presumably nonuniversal.
It turns out, however, that in this model we can estimate the
value of $b$ by an independent calculation.
In particular, simple qualitative arguments show that $b < 0$,
so that the asymptotic value of $\xi/e^{2\beta}$
should be reached from below as $\beta\to\infty$
(and likewise for the staggered susceptibility).

Such an increase, following a minimum, was in fact found in extensive
Monte Carlo simulations of this model by Ferreira and Sokal~\cite{FS}.
These authors attempted to fit their data with a variety of ``standard''
forms of corrections to scaling, including logarithms.
Though some of these Ans\"atze gave reasonable fits,
none had a plausible theoretical basis
(see \cite[Section 7.1]{FS} for detailed discussion).
As we shall show in Section~\ref{sec4}, the anomalous behavior
found by Ferreira and Sokal is consistent
with the unusual form (\ref{corr}) over its expected range of validity,
when $b$ is given the value that we extract independently
in Section~\ref{sec3}. 

Our ability to make such a theoretical analysis depends
on a mapping of the model at zero temperature
to a discrete height model that can be connected
(via a nonrigorous but convincing renormalization-group argument)
to the continuum theory of a free scalar boson
\cite{Burton_Henley_97,Henley_unpublished,Salas-Sokal_98}.
It is within the height-model approach
that we are able to disentangle the behavior at nonzero temperature
and argue that the two types of excitation,
leading to the unusual scaling form (\ref{corr}), are present.

The layout of this paper is as follows:
In Section~\ref{sec2} we discuss the mapping to a height model
and make the important observation that in this model
(unlike some other models with zero-temperature critical points)
the height mapping is well-defined also at nonzero temperatures.
It is then straightforward to identify the two types of excitation
in height-model language.  The most relevant excitation has the nature of a
vortex within the free-boson description, and standard Coulomb-gas
methods then lead to the prediction that $\nu=2$. The other type of
excitation has zero vorticity and, we argue, merely renormalizes the
compactification radius (or stiffness coefficient) of the free field.
However, this is not a rigorous argument,
and it is important to check it independently.
This we do in Section~\ref{sec3}, by a direct numerical investigation of the   
modified height model in which only excitations of the second type are allowed,
that is, the vortices are suppressed. In this case, the height model may
be represented as a modified vertex model, which we analyze using transfer
matrices and finite-size scaling. Our results show clearly that these
excitations are indeed strictly marginal: the modified vertex model is
still critical and continues to have the central charge $c=1$ characteristic
of a free boson. We directly measure the dependence of the
stiffness coefficient on the fugacity
$w \equiv e^{-\beta}$ of the excitations.
In Section~\ref{sec4} we put these pieces of information together
to predict how the effective critical exponents like $\nu_{\rm eff}$
depend on $e^{-\beta}$, and hence extract the value of the parameter $b$
in Eq.~(\ref{corr}) for both the correlation length and the susceptibility.
Finally we compare these predictions with the Monte Carlo data of
Ferreira and Sokal~\cite{FS}.
In the Appendix we show how various non-universal quantities arising
from the first-order effects of the non-vortex defects may be related to
one another.

\section{Height model and low-temperature excitations}
\label{sec2}

It is convenient to label the Potts states $s_i$ by the integers $0,1,2$.
The vertices $i$ of the square lattice are labeled by integer-valued
coordinates $(m_i,n_i)$, and the lattice is divided into even and odd
sublattices on which $m_i+n_i$ is even or odd, respectively.
Let us introduce the variable $\eta_i\equiv\frac12 [1-(-1)^{m_i+n_i}]$,
which takes the values 0 or 1 according to whether
$i$ is on the even or the odd sublattice.
We then define height variables $h_i \in {\bf Z}/6{\bf Z}$ by
\begin{equation}
\label{height}
h_i \;\equiv\; 2s_i+3\eta_i  \quad(\hbox{mod 6})   \;,
\end{equation}
as illustrated in the table below:
\begin{center}
\begin{tabular}{|c||c|c|}\hline
Potts state & Even Subl. & Odd Subl. \\ \hline
0           &  0         & 3         \\
1           &  2         & 5         \\
2           &  4         & 1         \\ \hline
\end{tabular}
\end{center}
This gives a 1--1 correspondence between configurations of the 3-state
Potts model and configurations of the height model satisfying
the constraint that $h_i$ is even (resp.\ odd) whenever $i$ is on the even
(resp.\ odd) sublattice.
This latter constraint may equivalently be imposed by fixing the height
at the origin to be even and further demanding that
heights on neighboring vertices $i$ and $j$ satisfy
\begin{equation}
 \label{eq2.2}
|h_i-h_j| \;=\; 1 {\rm\ or\ } 3   \quad(\hbox{mod 6})   \;.
\end{equation}
We then have $s_i \neq s_j$ if and only if $|h_i-h_j| = 1 \; (\hbox{mod 6})$,
and $s_i = s_j$ if and only if $|h_i-h_j| = 3 \; (\hbox{mod 6})$.

Let us consider first the model at zero temperature.
Then the only allowed configurations are antiferromagnetic ground states
($s_i \neq s_j$ for all nearest-neighbor pairs $i,j$);
in the height model this corresponds to replacing (\ref{eq2.2})
by the more restrictive condition 
\begin{equation}
\label{gs}
|h_i-h_j| \;=\; 1\quad(\hbox{mod 6})  \;.
\end{equation}
It then follows that the height field $h_i \in {\bf Z}/6{\bf Z}$
can be ``lifted'' to a height field $\widetilde{h}_i \in {\bf Z}$
so that now the height difference across an edge is
$\pm 1$ {\em tout court}\/ (not just mod 6):
self-consistency is ensured (at least with free boundary conditions)
by noting that
the change $\Delta \widetilde{h}$ around any plaquette will be zero
(if four numbers $\pm 1$ add up to 0 mod~6, they must necessarily be
   two $+1$'s and two $-1$'s, hence add up to 0).
Furthermore, this lifting is unique up to an overall shift by a multiple of 6.
In cylindrical or toroidal boundary conditions,
the height field $\widetilde{h}_i$ might fail to be globally well-defined
(i.e.\ it might have a nonzero tilt),
but the gradient field $\nabla \widetilde{h}$
is still well-defined and curl-free.

The restricted height model (\ref{gs}) can be equivalently be mapped onto
an arrow model on the dual lattice:
we assign to each edge $e$ of the dual lattice, dual to the edge $ij$
of the original lattice, the orientation obtained by a
$+ 90^\circ$ turn from the direction in which the height change is $+1$ (mod 6).
This arrow field is simply the dual of the vector field
$\nabla \widetilde{h}$;
it is therefore conserved at each vertex of the dual lattice.
Each vertex of the dual lattice thus has precisely two inward-pointing
arrows and two outward-pointing arrows,
so that the allowed configurations of the arrows at each vertex are
precisely those of the 6-vertex model \cite{Baxter_82}.
Moreover, the Boltzmann weights are those of the symmetric point,
i.e.\ all weights equal.
It is well known from the exact solution of the 6-vertex model
\cite{Baxter_82} that this point is critical.

As was pointed out by Henley \cite{Burton_Henley_97,Henley_unpublished},
this criticality is quite easily understood
from the point of view of the renormalization group
applied to the lifted height model.\footnote{
  One might ask why the RG has to be applied to the {\em lifted}\/
  (${\bf Z}$-valued) height model and not to the original
  (${\bf Z}/6{\bf Z}$-valued) height model.
  The answer is that the coarse-graining process necessarily introduces
  fractional weights,
  which would cause ambiguity in ${\bf R}/6{\bf Z}$
  but which give rise to a well-defined averaging operator in ${\bf R}$.
}
It is reasonable to guess that the long-wavelength behavior of the
lifted height model is controlled by an effective coarse-grained Hamiltonian
of the form
\begin{equation}
\label{SG}
S_{\rm G} \;=\; \int {\rm d}^2r \, \left[ {K \over 2} (\partial \widetilde{h})^2
                              \,-\, \lambda \cos(2\pi \widetilde{h}) \right]
\end{equation}
where $K$ is the stiffness constant. 
The gradient term in (\ref{SG}) takes into account
the entropy of small fluctuations around Henley's ``ideal states''
\cite{Kondev_96,Raghavan_97,Burton_Henley_97,Henley_unpublished,Salas-Sokal_98};
the second term is the so-called {\em locking potential}\/,
which favors the heights to take their values in ${\bf Z}$.
We then expect that there exists some constant $K_{\rm r}$ such that
for $K < K_{\rm r}$ (resp.\ $K > K_{\rm r}$) the locking potential is
irrelevant (resp.\ relevant) in the renormalization-group sense.
Thus, if $K < K_{\rm r}$ our surface model is ``rough'' and its
long-wavelength behavior can be described by a massless Gaussian model:
\begin{equation}
   \Bigl\langle [ \widetilde{h}(x) - \widetilde{h}(y)]^2
   \Bigr\rangle
   \;\sim\;  {1 \over \pi K} \, \log |x-y|
 \label{gaussian}
\end{equation}
for $|x-y|\gg 1$;
in this case, the original zero-temperature spin system is critical,
and all its critical exponents can be determined in terms of the
single constant $K$.
In particular, the scaling dimensions of spin-wave (or vertex) operators
$e^{i\alpha \widetilde{h}(r)}$ are given by
\begin{equation}
x_\alpha \;=\; \alpha^2/4\pi K.
\end{equation}
{}From (\ref{height}) the staggered order parameter $(-1)^{m_i+n_i}e^{2\pi
s_i/3}$ corresponds to $\alpha=\pi/3$, so that its correlation function
decays with an exponent $\eta_{\rm stagg}=2x_{\pi/3}=\pi/18K$.
By comparison with the exact result $\eta_{\rm stagg}=\frac13$
from the 6-vertex model \cite{Nijs_82,Park_89},
it follows that $K$ must take the value $\pi/6$.
At this value, the locking potential $\cos(2\pi \widetilde{h})$
has scaling dimension $x_{2\pi}=6$, so that it is indeed highly irrelevant.
By the usual scaling law we then obtain the susceptibility exponent
$(\gamma/\nu)_{\rm stagg} = 2 - \eta_{\rm stagg} = 5/3$.\footnote{
   This value has been confirmed numerically by several authors
   \cite{Wang_89,Wang_90,Salas-Sokal_98,FS}.
}

Let us now consider the model at nonzero temperature:
note that the height model (\ref{height})/(\ref{eq2.2})
continues to give a complete description.
Nonzero temperature amounts to allowing the height difference
between neighboring sites to take the value 3 (mod 6)
as well as $\pm 1$ (mod 6):
the former correspond to unsatisfied bonds.
Each unsatisfied bond will be weighted by $e^{-\beta}$,
so that, at low temperature, the unsatisfied bonds will be very dilute.
Let us therefore consider first the effect of just one
isolated unsatisfied bond.
This bond belongs to two neighboring plaquettes,
and there are two cases to consider,
according as the vorticities on those two plaquettes
have (a) opposite signs or (b) the same sign:
\begin{equation}
 \begin{array}{ccc}
   s \;=\;  \begin{array}{c@{\,}c@{\,}c}
               2 & & 1 \\
               0 & \hbox{---} & 0 \\
               2 & & 1 \\
            \end{array}
   & \qquad\qquad &
   s \;=\;  \begin{array}{c@{\,}c@{\,}c}
               2 & & 1 \\
               0 & \hbox{---} & 0 \\
               1 & & 2 \\
            \end{array}
   \\[8mm]
   {\rm (a)} & \qquad\qquad & {\rm (b)}
 \end{array}
\end{equation}
where the unsatisfied bond is indicated with a dash.
The corresponding height configurations mod 6 are
\begin{equation}
 \begin{array}{ccc}
   h \;=\;  \begin{array}{c@{\,}c@{\,}c}
               1 & & 2 \\
               0 & \hbox{---} & 3 \\
               1 & & 2 \\
            \end{array}
   & \qquad\qquad &
   h \;=\;  \begin{array}{c@{\,}c@{\,}c}
               1 & & 2 \\
               0 & \hbox{---} & 3 \\
               5 & & 4 \\
            \end{array}
   \\[8mm]
   {\rm (a)} & \qquad\qquad & {\rm (b)}
 \end{array}
 \label{eq.height}
\end{equation}
(here we have taken the center-left vertex to lie on the even sublattice).
Let us now ask whether these heights $h_i \in {\bf Z}/6{\bf Z}$
can be lifted to heights $\widetilde{h}_i \in {\bf Z}$
while satisfying the condition
\begin{equation}
|\widetilde{h}_i-\widetilde{h}_j| \;=\; 1 {\rm\ or\ } 3  \;.
\end{equation}
This amounts to asking whether the bond in (\ref{eq.height})
with $|h_i - h_j| = 3$
can be assigned a sign consistent with that of the two adjacent plaquettes.
In case (a), the answer is yes;
in case (b), the answer is no.
We therefore call situation (a) a {\em non-vortex unsatisfied bond}\/,
and situation (b) a {\em vortex}\/ (of strength 6).

This situation can alternatively be viewed in the 6-vertex picture
(Figure~\ref{fig1}).
The arrow dual to the unsatisfied bond may be thought of as an
arrow of strength 3, but of indeterminate sign. There are then
two possible ways in which this defect may be healed locally: either 
we can choose a definite orientation for the triple arrow,
and impose strict conservation at the neighboring vertices,
corresponding to the situation shown in Figure~\ref{fig1}a;
or we can relax this condition and allow a violation of arrow conservation
at one of the vertices, as in Figure~\ref{fig1}b.
In the latter case there is a net arrow flux of $\pm6$ 
out of the region of the defect, and this will persist to larger
distances if there is strict conservation elsewhere.

Clearly this kind of defect has a topological nature:
it corresponds to a vortex (or antivortex) in which
the height field changes by $\pm 6$ in encircling the defect. 
Outside the core of the defect, the continuum description (\ref{SG})
should still apply. Thus we may write
$\widetilde{h}(r) \approx (\pm6/2\pi)\theta+\widetilde{h}'$,
where $\theta$ is the polar angle and $\widetilde{h}'$ has zero vorticity.
Substituting into (\ref{SG}) we then find that the defect has additional
reduced free energy (i.e.\ negative entropy)
\begin{equation}
\label{vortex}
 \ffrac12K\left( {6 \over 2\pi} \right)^{\! 2}
    \int {\rm d}^2\!r/r^2  \;\sim\;  (9K/\pi)\log(R/a)
\end{equation}
where we have introduced a short-distance cutoff $a$, and $R$ is the
size of the system. Thus a single isolated vortex has zero probability of
occurring. However, configurations containing several vortices and antivortices
with zero total vorticity have a nonzero probability as $R\to\infty$.
The scaling dimension $x_{\rm V}$ of a vortex operator may be
determined \cite[p.~121]{JCbook}
from the power-law decay of the vortex-antivortex correlation
function (i.e.\ the partition function with a vortex-antivortex pair
introduced) or, more easily, directly from (\ref{vortex}) as
\begin{equation}
\label{xV}
x_{\rm V}=9K/\pi=\ffrac32
\end{equation}
at $K=\pi/6$. If the vortices are the most relevant perturbation at
nonzero temperature, which we shall argue is the case, then (\ref{xV})
leads to the prediction $\nu = 1/y_{\rm V} = 1/(2-x_{\rm V}) = 2$. 
Since the scaling variable $t$ is proportional to the fugacity
$e^{-\beta}$ of the vortex, we predict that the correlation length
diverges as $\xi\propto (e^{-\beta})^{-\nu}=e^{2\beta}$
as $\beta\to\infty$.

On the other hand, the non-vortex defect shown in Figure~\ref{fig1}a has finite
additional negative entropy. It may be viewed as a tightly bound pair
of a vortex and antivortex of strength $\pm 3$.
Thus, at large distances, it corresponds to a dipole height field 
\begin{equation}
\label{dipole}
h(r)\sim D{{\bf n}\cdot{\bf r}\over r^2}
\end{equation}
where ${\bf n}$ is a unit lattice vector, and $D$ is a non-universal
constant whose value is difficult to determine analytically, since the
size of the vortex-antivortex pair is of the order of the lattice
spacing, at which scale the continuum action~(\ref{SG}) is inapplicable.
(Nevertheless, in the Appendix we show how the value of $D$ may be
related to other non-universal numbers that we have measured directly.)
The dipole field (\ref{dipole}) gives an infrared-finite
but nonzero contribution to $S_{\rm G}$
(of course the integral should still be cut off at short distances:
 it diverges as $a^{-2}$ and so is strongly dependent on the precise form
 of the cutoff).
The numerical value of this nonuniversal constant will be
measured in Section~\ref{sec3}.

As already noted, the non-vortex defect has the property
--- at least in the simple case when it is isolated ---
that it is possible to assign a definite value $+3$ or $-3$
to the height change $\Delta \widetilde{h}$ along the defect edge
in such a way that the gradient field $\nabla \widetilde{h}$
remains curl-free, i.e.\ such that the height variables
are lifted locally to ${\bf Z}$.
(In the 6-vertex picture, this means that it is possible to assign
a definite arrow value $+3$ or $-3$ to the defect edge in such a way that
strict conservation holds.)
Even when the defects are not isolated, but form finite clusters,
as long as there is no net flux of arrows from each cluster,
it should still be possible to lift the height variables locally to ${\bf Z}$
in a unique manner, except possibly in the core of the cluster.
(An example of ambiguity within the core of the cluster
 is shown in Figure~\ref{fig2}.)
We therefore expect that non-vortex defects do not disturb the
renormalization-group flow of the model towards a free boson theory
as in (\ref{SG}).  On the other hand, there is no reason to suppose
that a model with a finite density of defects will renormalize onto
the {\em same value}\/ of the stiffness constant $K$ as is obtained
in the absence of defects;  rather, we expect that $K$ will depend on
the fugacity $w = e^{-\beta}$ of the non-vortex defects.
It is the purpose of the next section to test this hypothesis numerically
and to estimate the dependence $K_{\rm eff}(w)$
in the model with vortices suppressed.


In general, the renormalization-group flows in a model with a scale-dependent
stiffness constant $K(\ell)$ and vortex fugacity $y(\ell)$ are
well-known~\cite{Kosterlitz} to be
(for a vortex of strength 6)
\begin{subeqnarray}
{\rm d}y/{\rm d}\ell&=&(2-9K/\pi)y+O(y^3)  \slabel{rg1}\\
{\rm d}K^{-1}/{\rm d}\ell&=&Ay^2+O(y^4)    \slabel{rg2}  \label{rg1+2}
\end{subeqnarray}
The coefficient of $y$ in the first equation
reflects the fact that our vortices have scaling dimension $x_{\rm V}=9K/\pi$.
The coefficient $A$ in the second equation is nonuniversal
(since it depends on the normalization of $y$)
but of order unity;
it is independent of $K$ because
the vortex-vortex interaction is $\propto K$,
and the screening term from tightly bound pairs is $\propto K^2$,
so that ${\rm d}K/{\rm d}\ell \propto K^2 y^2$.

Defining $u=\pi/6-K$, the RG flow (\ref{rg1+2}) may be rewritten as
\begin{subeqnarray}
{\rm d}y/{\rm d}\ell&=&(\ffrac12+9u/\pi)y+O(y^3)   \slabel{rg3}\\
{\rm d}u/{\rm d}\ell&=&A'y^2+O(y^4, y^2 u)         \slabel{rg4}
\end{subeqnarray}
where $A'=(\pi/6)^2A$. Away from the point where $u = -\pi/18$ (where the
vortices become marginal), these equations may be rewritten in the
standard way \cite{wegner} in terms of non-linear scaling variables
$\tilde u=u+O(y^2)$ and $\tilde y=y+O(y^3)$ so that the flow equations
take the simple form
\begin{subeqnarray}
{\rm d}\tilde y/{\rm d}\ell&=&(\ffrac12+9\tilde u/\pi)\tilde y \slabel{rg5}\\
{\rm d}\tilde u/{\rm d}\ell&=&0                                \slabel{rg6}
\end{subeqnarray}
with no higher-order terms. 
These equations are to be integrated with the initial conditions
$y(0) = w = e^{-\beta}$ and $K(0) = K_{\rm eff}(w)$,
hence
$\tilde y(0)=y(0)+O(y(0)^3)=w+O(w^3)$ and 
$\tilde u(0)=u(0)+O(y(0)^2)=\frac{\pi}6-K_{\rm eff}(w)+O(w^2)
            = - K'_{\rm eff}(0) w + O(w^2)$.
Thus $\tilde u(\ell)=\tilde u(0)$ and
\begin{equation}
\tilde y(\ell) \;=\; [w+O(w^3)] \,
                     \exp\left[(\ffrac12+9\tilde u(0)/\pi)\ell \right]
   \;.
\end{equation}

The correlation length satisfies the homogeneous renormalization-group
equation
\begin{equation}
\xi(\tilde u(0),\tilde y(0)) \;=\; e^\ell \, \xi(\tilde u(\ell),\tilde y(\ell))
   \;.
\end{equation}
Assuming in the standard way that $\xi=\xi_0=O(1)$ when $y(\ell)=O(1)$
[i.e., when $\tilde y(\ell)=O(1)$] then gives the prediction
\begin{subeqnarray}
\xi(\beta)&\sim&\xi_0 \, \big[w+O(w^3)\big]^{-1/(\frac12+9\tilde{u}/\pi)} \\
  &\sim&\xi_0 \, \big[w+O(w^3)\big]^{-[2+(36/\pi)K_{\rm eff}'(0)w+O(w^2)]}\\
  &=&  \xi_0 \, \exp\!\left[ 2\beta + b\beta e^{-\beta} + O(\beta e^{-2\beta})
      \right] \\
&=&\xi_0 \, e^{2\beta} \left[1+b\beta e^{-\beta}+\ffrac12b^2\beta^2e^{-2\beta}
                              + O(\beta e^{-2\beta}) \right]
\end{subeqnarray}
where $b=(36/\pi)K_{\rm eff}'(0)$. In writing the above we have been
careful to show where the neglected higher-order terms enter.
Simple qualitative arguments (see Section~\ref{sec4})
show that $K_{\rm eff}'(0)<0$,
so this implies that
the asymptotic value of $\xi(\beta)/e^{2\beta}$ is attained from below.

Next we turn to the staggered susceptibility.
A staggered field $h_{\rm stagg}$ satisfies the renormalization-group equation
\begin{equation}
   {\rm d}h_{\rm stagg}/{\rm d}\ell  \;=\;
      (2-x_{\pi/3})h_{\rm stagg}+O(h_{\rm stagg}^3)
\end{equation}
where, as discussed above, $x_{\pi/3}=\pi/36K$.
As usual, the singular part of the reduced free energy per unit area
transforms according to $f(\tilde u(0),\tilde y(0),h_{\rm stagg}(0))=
e^{-2\ell}f(\tilde u(\ell),\tilde y(\ell),h_{\rm stagg}(\ell))$, so that
the staggered susceptibility 
$\chi_{\rm stagg}\sim\partial^2f/\partial h_{\rm stagg}^2$ satisfies
\begin{equation}
\chi_{\rm stagg}(\tilde u(0),\tilde y(0))   \;=\;
  \exp\!\left[ \int_0^\ell \, [2-2x_{\pi/3}(\ell')] \, d\ell' \right]
   \chi_{\rm stagg}(\tilde u(\ell),\tilde y(\ell))
\end{equation}
where the integral in the exponential is
\begin{equation}
\int_0^\ell\big[\ffrac53-(2\tilde u/\pi)+O(\tilde y(\ell')^2)\big] \, d\ell'
   \;\sim\; (\ffrac53-2\tilde{u}/\pi)\ell+{\rm const}+O(w^2)
   \;.
\end{equation}
Choosing $\ell$ as before and assuming that at this scale
$\chi_{\rm stagg}=\chi_0=O(1)$ then gives
\begin{equation}
\label{chi1}
\chi_{\rm stagg} \;\sim\; \chi_0 \,
\big[w+O(w^2)\big]^{-(\frac53-2\tilde{u}/\pi)/(\frac12+9\tilde{u}/\pi)+O(w^2)}
\end{equation}
where, once again, $\tilde{u}=-K_{\rm eff}'(0)w + O(w^2)$.
Equation (\ref{chi1}) then simplifies to the final result
\begin{subeqnarray}
\chi_{\rm stagg}(\beta) &\sim&
 \chi_0 \, \exp\!\left[ \ffrac{10}{3} \beta + b'\beta e^{-\beta}
                         + O(\beta e^{-2\beta})\right]  \\
 &=&  \chi_0 \,
e^{{10 \over 3}\beta} \left[1+b'\beta e^{-\beta}+\ffrac12{b'}^2 \beta^2
e^{-2\beta} +O(\beta e^{-2\beta})\right]
\end{subeqnarray}
where $b'=(64/\pi)K_{\rm eff}'(0)=\frac{16}{9}b$.
Thus the asymptotic limit of
$\chi_{\rm stagg}(\beta)/e^{{10 \over 3}\beta}$ is also attained from below.

\section{Extracting the stiffness constant}
\label{sec3}

In this section we analyze a model in which the vortex excitations present
in the full model are deliberately suppressed,
in order to understand the effects of the (less relevant) non-vortex defects
like those shown in Figure~\ref{fig1}a.
As discussed in the previous section, this may be achieved in the
height model by allowing height changes $\Delta \widetilde{h}$
of $\pm 3$ as well as $\pm 1$,
while continuing to insist that the gradient field $\nabla \widetilde{h}$
be strictly curl-free.
Edges with $\Delta \widetilde{h} = \pm 3$ will be assigned
a fugacity $w \equiv e^{-\beta}$.
Equivalently, we may work in the dual arrow model:
edges in this model are allowed to carry flux $\pm 3$ as well as $\pm 1$,
and we impose strict flux conservation at the vertices.
The partition function of the arrow model is therefore
\begin{equation}
 Z \;=\; \sum_{\cal G} w^{N_3}  \;,
 \label{Zw}
\end{equation}
where $N_3$ is the number of triple arrows,
and the sum $\sum_{\cal G}$ extends over all possible flux-conserving
configurations of the 44-vertex model just defined.\footnote{
   Barbero {\em et al.}\/ \cite{Barbero_97} recently studied
   a closely related 14-vertex model, in which zero or one edges
   of strength $\pm 3$ (but {\em not}\/ two or four such edges)
   are allowed to be incident on each vertex.
   For small fugacity $w$, this model should be essentially
   equivalent to our model;
   in particular, we expect it to remain in the ``rough'' phase
   for small $w$,
   with a continuum limit given by the massless free field (\ref{SG})
   with a stiffness constant $K$ that depends on $w$.
   However, this model cannot stay critical for all $w > 0$,
   since in the limit $w \to\infty$ it freezes into one of a finite
   number of ground states
   (for example, vertices 11 and 16 of \cite[Figure 1]{Barbero_97}
    on even and odd sublattices, respectively).
   It would be interesting to understand what happens in-between.
   Unfortunately, the study of Barbero {\em et al.}\/ is limited
   to the region $0 \le w \le 1$
   (corresponding to $\beta \epsilon_3 \ge 0$ in their notation;
    we are at the symmetric point $\beta \epsilon = 0$).
   We thank Giorgio Mazzeo for bringing this paper to our attention.
}
We wish to test the hypothesis that the
continuum limit of the statistical model defined by (\ref{Zw}) is
a free-field theory with action given by (\ref{SG}),
with a stiffness constant $K$ that depends on $w$.
In this section we shall see how it is possible to verify this hypothesis
numerically, and at the same time extract quite accurate values of 
$K_{\rm eff}(w)$
as well as its first few derivatives at $w=0$.

Imagine defining the model (\ref{Zw}) on a cylinder of circumference $L$,
with periodic boundary
conditions in the transverse direction. Clearly, due to the flux conservation
it splits up in a direct sum of theories with a fixed net arrow current $Q$
in the longitudinal direction.
The corresponding height field
$\widetilde{h}({\rm r}) \equiv \widetilde{h}(x,t)$
is multiple-valued in the transverse coordinate $x$,
with ``tilt'' $Q$:  $\widetilde{h}(x+L,t) = \widetilde{h}(x,t) + Q$. 
Such a constant tilt can, however, easily be gauged away by setting
\begin{equation}
 \widetilde{h}(x,t)  \;=\;  Qx/L + \widehat{h}(x,t)  \;,
 \label{shift}
\end{equation}
where now $\widehat{h}(x+L,t) = \widehat{h}(x,t)$.
For $Q/L \ll 1$ the field $\widehat{h}$ can be assumed to describe the $Q=0$
sector of the original model. Therefore, inserting (\ref{shift}) into
(\ref{SG}), we see that a nonzero current $Q$ simply shifts the
action (free energy) per unit area by an amount
\begin{equation}
 \Delta f(Q) \;=\; \frac{K_{\rm eff}(w)}{2} \, \frac{Q^2}{L^2}  \;.
 \label{f(Q)}
\end{equation}
This is in turn related to the shift in the logarithm of the largest
eigenvalue of the corresponding transfer matrix, which can then be used to
check the above hypothesis and to extract 
numerical values of $K_{\rm eff}(w)$.

To this end we have constructed the transfer matrix of the model $(\ref{Zw})$
in the
$4^L$-dimensional basis associated with all possible arrow configurations
for a layer of $L$ vertical bonds. As mentioned above, the transfer matrix is
block-diagonal according to the value of $Q$. Furthermore, $Q$ must have the
same parity as $L$. Using standard sparse-matrix techniques we have been
able to diagonalize the various sectors of this matrix for
widths up to $L_{\rm max}=10$. Due to parity effects we limit the discussion
to even $L$ in the following, although odd $L$ yield compatible results.

To check that our model is indeed described by a Gaussian theory for all
$w \ge 0$, we begin by examining the central charge. This can be extracted
from the finite-size scaling of the free energy per unit area \cite{Cardy86}
\begin{equation}
  f_0(L) = f_0(\infty) - \frac{\pi c}{6 L^2} + \cdots  \;,
  \label{f0}
\end{equation}
where $f_0(L) = -\frac{1}{L} \log \lambda_0(L)$ and $\lambda_0(L)$ is the
largest eigenvalue of the transfer matrix in the $Q=0$ sector.
As usual, the convergence can be sped up
by including a nonuniversal $1/L^4$ correction, so that three
consecutive system sizes are needed to fit the above formula \cite{Cardy98}.
The resulting finite-size estimates for $c$ as a function of $w$ are shown
in Table~\ref{tab1}. As expected, they strongly suggest the $w$-independent
value $c=1$ in the $L \to \infty$ limit, thus corroborating the analytical
arguments given in Section~\ref{sec2}. (For reasons that we do not fully
understand, the finite-size effects are particularly strong at $w \approx 2$.
This would be an interesting question to investigate further.)

\begin{table}
 \begin{center}
 \begin{tabular}{r|ccc}
   $w$ & $c(6)$ & $c(8)$ & $c(10)$ \\ \hline
   0.0 & 0.9751 & 0.9885 & 0.9958  \\
   0.5 & 1.0079 & 0.9971 & 0.9990  \\
   1.0 & 0.9430 & 0.9812 & 0.9968  \\
   1.5 & 1.3336 & 1.0448 & 0.9407  \\
   2.0 & 1.6105 & 1.3984 & 1.2046  \\
   3.0 & 0.9486 & 0.9335 & 0.9771  \\
  10.0 & 0.9618 & 0.9882 &         \\
 \end{tabular}
 \end{center}
 \protect\caption[2]{\label{tab1}Effective central charge $c$ as a function
 of the Boltzmann weight $w$. The three-point fits to (\ref{f0}) based
 on system sizes $L$, $L-2$ and $L-4$ are labeled as $c(L)$.}
\end{table}

We now turn our attention to the extraction of the stiffness constant 
$K_{\rm eff}(w)$.
For each $L \le L_{\rm max}$ we have verified that the $Q$-dependence
of the free energy is indeed of the form (\ref{f(Q)}) for $Q \ll L$.
However, as the quadratic behavior must (and does) saturate
when $Q$ becomes comparable to $L$, we have
based our estimates of $K_{\rm eff}(w)$ on only the two lowest permissible values of $Q$.
For even $L$ these values must be even, due to the parity observation
made above;  this suggests that we focus on $Q=0$ and $Q=2$.
But at this point special attention must be paid to the limit $w \to \infty$,
in which only triple arrows are allowed, so that the model reduces
to the usual 6-vertex model with triple-size heights.
Such configurations are incompatible with a total flux of $Q=2$.
Thus, for generic (i.e.\ non-small) values of $w$ we should restrict attention
to values of $Q$ that are multiples of 6, e.g.~$Q=0$ and $Q=6$.

\begin{table}
 \begin{center}
 \begin{tabular}{r|cccl}
   $w$ & $K(L=6)$ & $K(L=8)$ & $K(L=10)$ & Extrapolation\\ \hline
   0.0 & 0.505099 & 0.512823 & 0.516574  & $0.52359 (1) \approx \pi/6.000$ \\
   0.5 & 0.207033 & 0.205920 & 0.205408  & $0.20450 (1) \approx \pi/15.4$ \\
   1.0 & 0.125262 & 0.125426 & 0.125496  & $0.12562 (2) \approx \pi/25.0$ \\
   1.5 & 0.091195 & 0.092921 & 0.093571  & $0.0947 (4)  \approx \pi/33.2$ \\
   2.0 & 0.064168 & 0.064762 & 0.064972  & $0.0654 (2)  \approx \pi/48.1$ \\
   3.0 & 0.057232 & 0.058096 & 0.058511  & $0.05925 (5) \approx \pi/53.0$ \\
 \end{tabular}
 \end{center}
 \protect\caption[2]{\label{tab2}Finite-size estimates for the stiffness
 constant $K_{\rm eff}(w)$ as well as their extrapolation to $L=\infty$.}
\end{table}

The values of $K_{\rm eff}(w)$ 
extracted by fitting (\ref{f(Q)}) at $Q=0$ and $Q=6$
are displayed in Table~\ref{tab2} for selected values of $w$ and $L$.
We observe a leading finite-size correction in $1/L^2$,
reflecting the $1/L^4$ correction present in $c$.
Thus, fitting each pair of successive system sizes to the form
$K_{\rm eff}(L) = K_{\rm eff}(\infty) - \mbox{\rm const}/L^2$,
we obtain two independent estimates of $K_{\rm eff}(\infty)$;
and judging the error bar from the small residual size dependence
of these estimates we obtain the final result shown in the rightmost
column of Table~\ref{tab2}.
For $w=0$ we find excellent agreement with the exact
result for the 6-vertex model, $K_{\rm eff}(0) = \pi/6$.
For $w \to\infty$ the triple-height model should have correlations
(\ref{gaussian}) nine times as large, hence a stiffness constant
one-ninth as large, i.e.\ $K_{\rm eff}(\infty)=\pi/54$;
this too is confirmed numerically.
In-between, $K_{\rm eff}(w)$ is a 
monotonically decreasing function of $w$;
this makes sense heuristically, because
allowing $\Delta\widetilde{h}=\pm3$ increases the variance of the
distribution of nearest-neighbor height differences,
which should lead (or so one naturally expects) to a larger variance
also for the long-distance height differences,
which by (\ref{gaussian}) are proportional to $K^{-1}$. 

\begin{table}
 \begin{center}
 \begin{tabular}{r|ccccl}
   Quantity & $L=4$     & $L=6$     & $L=8$     & $L=10$ &
                                                  Extrapolation\\ \hline
 $K'(0)$  & $-0.330292$ & $-0.452044$ & $-0.502110$ & $-0.527432$ & $-0.58$ (1)
   \\
 $K''(0)$ & $-3.10237$  & $-2.69860$  & $-2.49302$  & $-2.37655$  & $-2.1$ (1)
   \\
 \end{tabular}
 \end{center}
 \protect\caption[2]{\label{tab3}Finite-size estimates and extrapolations
 for the derivatives $\left.{\rm d}K/{\rm d}w\right|_{w=0}$ and 
 $\left.{\rm d}^2K/{\rm d}w^2\right|_{w=0}$.}
\end{table}

Exactly at $w=0$ the above remark that the $Q=2$ sector introduces frustration
does not apply. The evaluation of $K$ and its first few derivatives at $w=0$ 
can therefore be based on the $Q=0$ and $Q=2$ sectors, yielding a better
precision. To perform the derivatives, we simply numerically differentiate
the free energies (\ref{f(Q)}).
The results for $K'(0)$ and $K''(0)$ are given in Table~\ref{tab3}.

Our transfer-matrix results can also be used to compute the average energy
density, given here simply as the probability of having a triple arrow,
$e(w) = \langle N_3 \rangle/N$.
Using (\ref{Zw}) we have $e(w) = -w \, df/dw$, and 
based on the first two derivatives of the free energy we find
\begin{equation}
\label{energy2}
 e(w) \;=\; 0.2179(2) w + 0.696(1) w^2 +  O(w^3)
\end{equation}
in the model with vortices suppressed.
This can be compared with the Monte Carlo results of Ferreira and Sokal
for the full 3-state Potts antiferromagnet (i.e.\ with vortices included),
based on a fit to the low-temperature data
\cite[Section 4.3 and Figure 13]{FS}:
\begin{equation}
\label{energy1}
 e_{\rm Potts}(w) \;=\; 0.21777 w + 1.65303 w^2 +  O(w^3)   \;.
\end{equation}
The fact that the leading terms of the two expansions agree could easily
have been anticipated, since vortices necessarily come in pairs, and thus
enter only at order $O(w^2)$.

\section{Comparison with Monte Carlo results}
\label{sec4}

Putting together the results of Sections~\ref{sec2} and \ref{sec3},
we deduce the theoretical predictions
\begin{eqnarray}
   \log \xi(\beta)   & = &  2\beta \,+\, \log\xi_0 \,+\, b \beta e^{-\beta}
                           \,+\, O(\beta e^{-2\beta})  \label{theor_xi} \\[2mm]
   \log \chi(\beta)  & = &  \ffrac{10}{3}\beta \,+\, \log\chi_0 \,+\,
                               b' \beta e^{-\beta} \,+\, O(\beta e^{-2\beta})
                                                       \label{theor_chi}
\end{eqnarray}
where the nonuniversal constants $b$ and $b'$ take the values
\begin{eqnarray}
   b   & = &  {36 \over \pi} K'_{\rm eff}(0)  \;\approx\;  -6.65(11)
     \label{theor_b}   \\
   b'  & = &  {64 \over \pi} K'_{\rm eff}(0)  \;\approx\;  -11.82(20)
     \label{theor_bprime}
\end{eqnarray}
and we write $\chi$ as a shorthand for $\chi_{\rm stagg}$.
We are now ready to compare these predictions
with the Monte Carlo data of Ferreira and Sokal
\cite[Table 4, $L_{min} = 128$]{FS}.

Let us begin with the correlation length.
In Figure~\ref{fig.sec4.plot1} we plot $\xi(\beta)/e^{2\beta}$ versus $\beta$;
the rise at $\beta \gtapprox 3.4$ (corresponding to $\xi \gtapprox 75$)
is seen clearly.
In Figure~\ref{fig.sec4.plot2} we replot the same data
as $\log \xi(\beta) - 2\beta$ versus $\beta e^{-\beta}$;
there is a lot of curvature, and it is unfeasible to extract
reliable estimates of the limiting slope,
but the data for $\beta e^{-\beta} \ltapprox 0.031$
(corresponding to $\beta \gtapprox 5.1$ and $\xi \gtapprox 2600$)
are at least {\em compatible}\/ with the predicted slope $-6.65$,
provided that we choose an intercept $\log\xi_0 = -2.11(2)$.
In Figure~\ref{fig.sec4.plot3} we plot
$\log \xi(\beta) - 2\beta + 6.65 \beta e^{-\beta}$ versus $\beta e^{-\beta}$;
for modest values of $\beta$ the corrections to scaling are actually
{\em stronger}\/ than in Figure~\ref{fig.sec4.plot2} (!),
but for large $\beta$ they are weaker and it is at least plausible
that the curve is asymptotically horizontal.
Finally, in Figure~\ref{fig.sec4.plot4} we plot
$\log \xi(\beta) - 2\beta + 6.65 \beta e^{-\beta}$ versus $\beta e^{-2\beta}$;
now the curve is compatible with linearity over the much wider range
$\beta e^{-2\beta} \ltapprox 0.0019$
(corresponding to $\beta \gtapprox 3.8$ or $\xi \gtapprox 170$),
and we estimate an intercept $\log\xi_0 = -2.12(1)$
and an asymptotic slope $C = 123(4)$.

Next we analyze the staggered susceptibility.
We omit the plot of $\chi(\beta)/e^{(10/3)\beta}$ versus $\beta$;
it looks a lot like Figure~\ref{fig.sec4.plot1}.
In Figure~\ref{fig.sec4.chiplot2} we plot
$\log \chi(\beta) - \ffrac{10}{3}\beta$ versus $\beta e^{-\beta}$;
once again there is a lot of curvature, and it is unfeasible to extract
reliable estimates of the limiting slope,
but the data for $\beta e^{-\beta} \ltapprox 0.031$
are {\em compatible}\/ with the predicted slope $-11.82$,
provided that we choose an intercept $\log\chi_0 = -2.52(3)$.
In Figure~\ref{fig.sec4.chiplot3} we plot
$\log \chi(\beta) - \ffrac{10}{3}\beta + 11.82 \beta e^{-\beta}$
versus $\beta e^{-\beta}$;
for modest values of $\beta$ the corrections to scaling are once again
{\em stronger}\/ than in Figure~\ref{fig.sec4.chiplot2},
but for large $\beta$ they are weaker and it is at least plausible
that the curve is asymptotically horizontal.
Finally, in Figure~\ref{fig.sec4.chiplot4} we plot
$\log \chi(\beta) - \ffrac{10}{3}\beta + 11.82 \beta e^{-\beta}$
versus $\beta e^{-2\beta}$;
now the curve is compatible with linearity over the much wider range
$\beta e^{-2\beta} \ltapprox 0.0016$
(corresponding to $\beta \gtapprox 3.9$ or $\xi \gtapprox 210$),
and we estimate an intercept $\log\chi_0 = -2.54(1)$
and an asymptotic slope $C' = 232(8)$.

In conclusion, the available Monte Carlo data are {\em compatible}\/
with the theoretical predictions (\ref{theor_xi})/(\ref{theor_chi}),
although the evidence for these predictions
--- and in particular for the predicted values
(\ref{theor_b})/(\ref{theor_bprime})
of the nonuniversal constants $b$ and $b'$ ---
is admittedly less than overwhelming.
Our analysis does, in any case, give a simple explanation
of why the limiting values of
$\xi(\beta)/e^{2\beta}$ and $\chi(\beta)/e^{(10/3)\beta}$
are approached from below.
It would be useful to obtain higher-precision Monte Carlo data
at correlation lengths $\xi \gtapprox 1000$
(corresponding to $\beta \gtapprox 4.6$)
in order to make a better test of our theoretical predictions.
It is curious that the asymptotic behavior in this model
is attained only at rather large correlation lengths.

\section*{Acknowledgments}

We wish to thank Marcel den Nijs and Sergio de Queiroz
for helpful conversations.
We also wish to thank Chris Henley for helpful correspondence
with one of us (A.D.S.)\ in 1994--98;
Henley was the first to notice the importance in this model
of vortices of strength $\pm 6$,
implying a leading behavior $\xi \propto e^{2\beta}$.

This research was supported in part by EPSRC grant GR/J 78327 (J.L.C.)\ 
and NSF grant PHY--9900769 (A.D.S.).
It was begun while A.D.S.~was a
Visiting Fellow at All Souls College, Oxford,
where his work was supported in part by EPSRC grant GR/M 71626
and aided by the warm hospitality of the Department of Theoretical Physics.
It was continued while J.L.J.~and J.L.C.~were hosted by the
Centre de Recherches Math\'ematiques, Montr\'eal, in connection with the
Concentration Period on ``Integrable models in condensed matter and
non-equilibrium physics''.

\appendix

\section{Relations between non-universal constants}

In this appendix we show how $\left. {\rm d}K/{\rm d}w \right|_{w=0}$
is related to two other non-universal constants:
$D$ in (\ref{dipole}) and
$c_1= \left. {\rm d}e/{\rm d}w \right|_{w=0}$
in (\ref{energy2})/(\ref{energy1}).

\begin{figure}
\centerline{
\epsfxsize=5in
\epsfbox{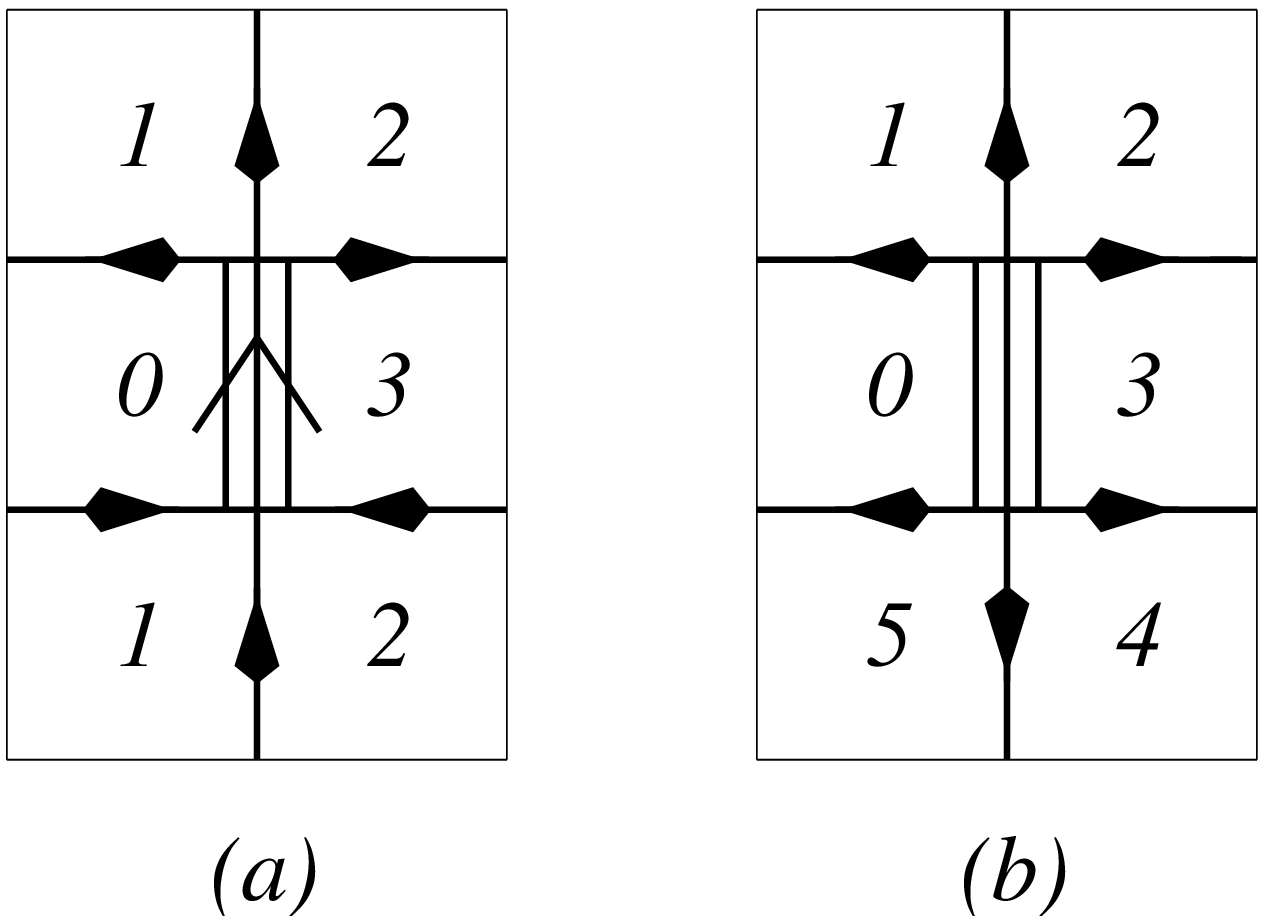}}
\caption{
   Examples of the two types of defect:
   (a) non-vortex unsatisfied bond, and (b) vortex.
   In both cases, two neighboring sites (labeled here by heights 0 and 3)
   are in the same Potts state (here $s=0$).
   Case $(a)$ is a local defect with finite relative entropy,
   while $(b)$ corresponds to a vortex in the 6-vertex model,
   with logarithmically diverging negative entropy.
}
\label{fig1}
\end{figure}

\begin{figure}
\centerline{
\epsfxsize=2.5in
\epsfbox{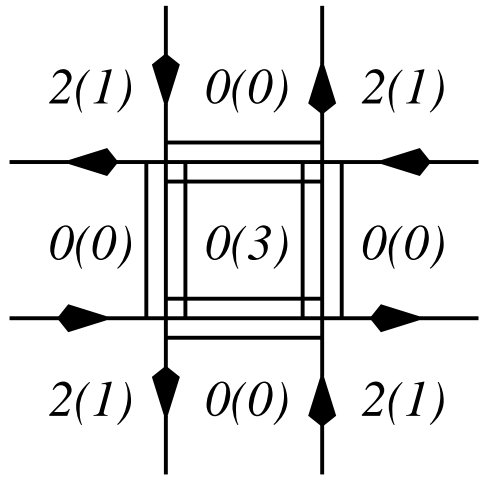}}
\caption{
   An extended non-vortex defect that corresponds to a well-defined set
   of Potts states (first label) but for which the heights mod 6
   (second label, in parentheses) cannot be lifted uniquely everywhere
   from ${\bf Z}/6{\bf Z}$ to ${\bf Z}$,
   since the height at the center could correspond to either $+3$ or $-3$.
   Equivalently,
   there are two flux-conserving ways of assigning triple arrows
   to the unsatisfied bonds (encircling the center plaquette clockwise
   or anticlockwise).
   Nevertheless, sufficiently far away from the defect there is no such
   ambiguity in lifting to ${\bf Z}$.
}
\label{fig2}
\end{figure}

\begin{figure}[p]
\centerline{
\epsfxsize=4in
\epsfbox{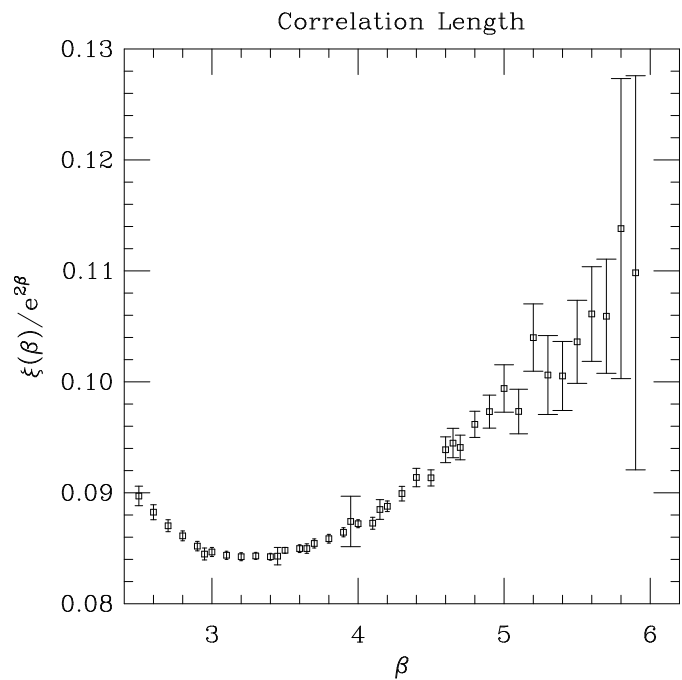}}
\vspace*{-2cm}
\caption{
   $\xi(\beta)/e^{2\beta}$ versus $\beta$.
   Error bars are one standard deviation,
   and are most likely overestimates \cite[Section 4.1.1]{FS}.
}
\label{fig.sec4.plot1}
\end{figure}

\begin{figure}[p]
\centerline{
\epsfxsize=4in
\epsfbox{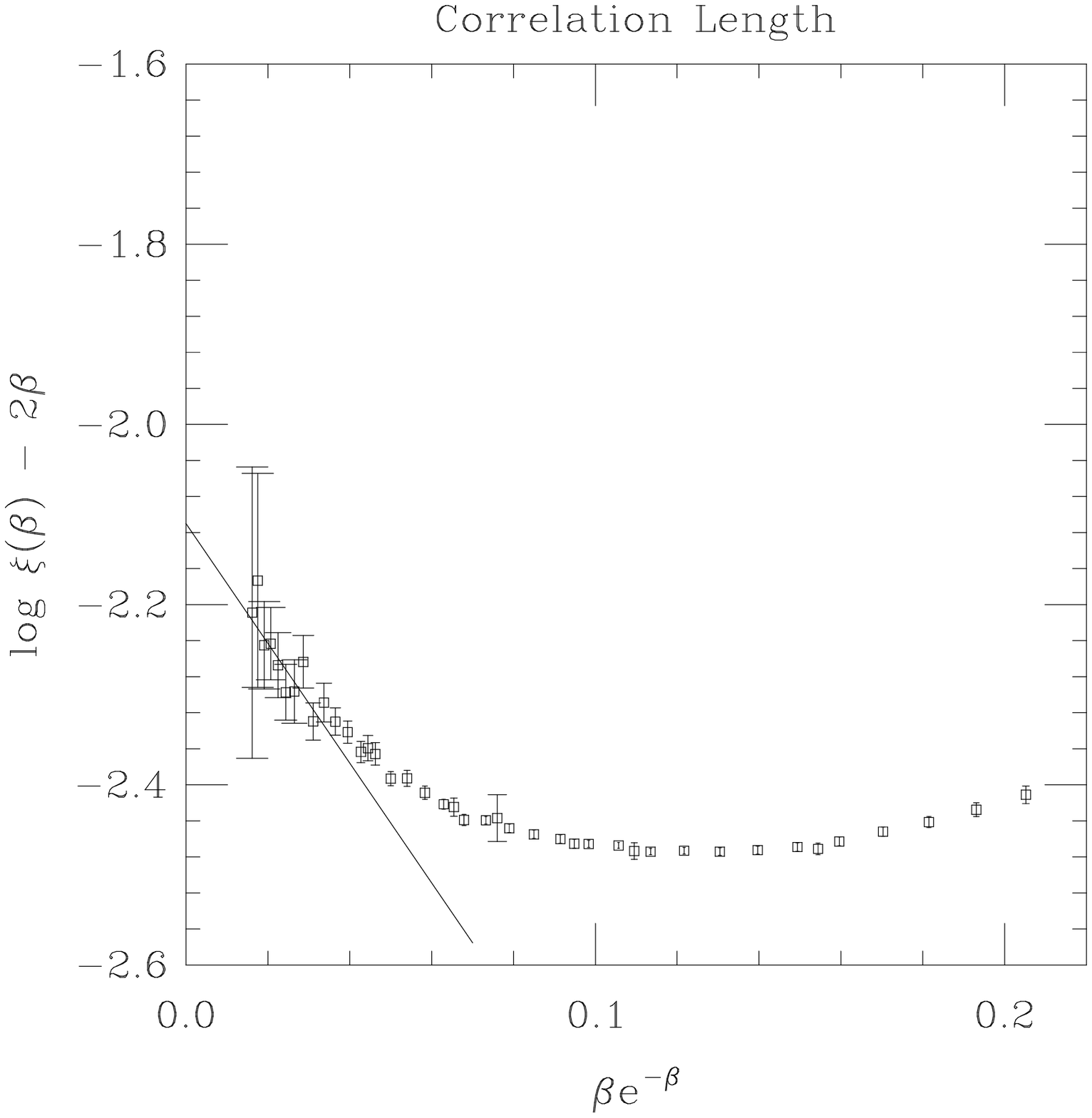}}
\vspace*{-1.5cm}
\caption{
   $\log \xi(\beta) - 2\beta$ versus $\beta e^{-\beta}$.
   Straight line is
   $\log \xi(\beta) - 2\beta = -2.11 - 6.65\beta e^{-\beta}$.
}
\label{fig.sec4.plot2}
\end{figure}

\begin{figure}[p]
\centerline{
\epsfxsize=4in
\epsfbox{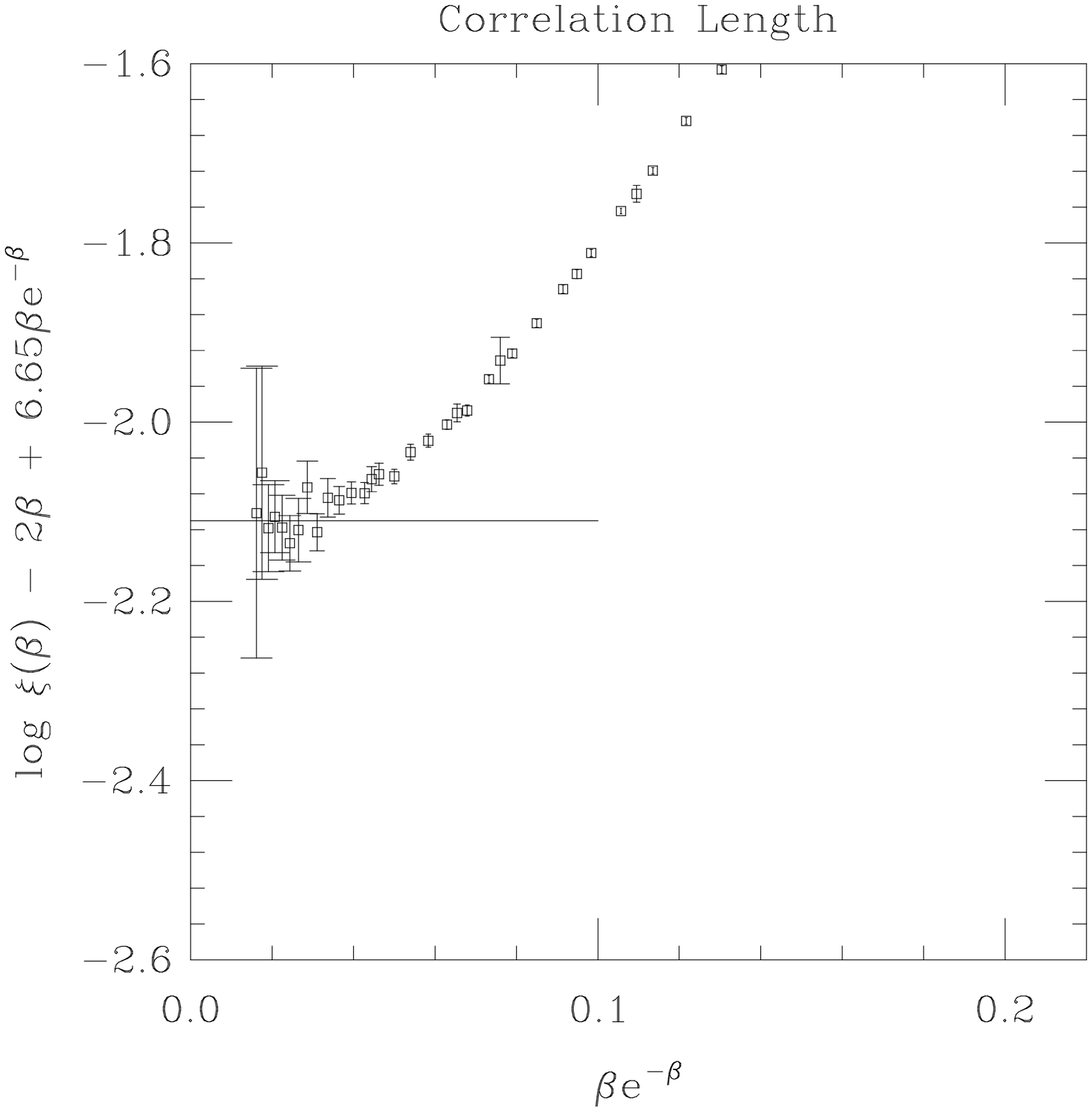}}
\vspace*{-1.5cm}
\caption{
   $\log \xi(\beta) - 2\beta + 6.65\beta e^{-\beta}$
   versus $\beta e^{-\beta}$.
   Straight line is
   $\log \xi(\beta) - 2\beta + 6.65\beta e^{-\beta} = -2.11$.
}
\label{fig.sec4.plot3}
\end{figure}

\begin{figure}[p]
\centerline{
\epsfxsize=4in
\epsfbox{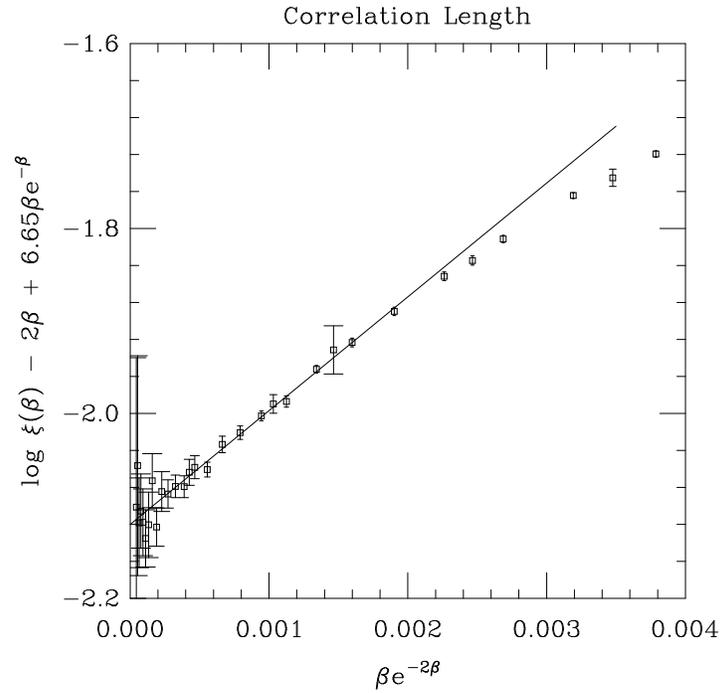}}
\vspace*{-1.5cm}
\caption{
   $\log \xi(\beta) - 2\beta + 6.65\beta e^{-\beta}$
   versus $\beta e^{-2\beta}$.
   Straight line is
   $\log \xi(\beta) - 2\beta + 6.65\beta e^{-\beta} =
    -2.12 + 123 \beta e^{-2\beta}$.
}
\label{fig.sec4.plot4}
\end{figure}

\begin{figure}[p]
\centerline{
\epsfxsize=4in
\epsfbox{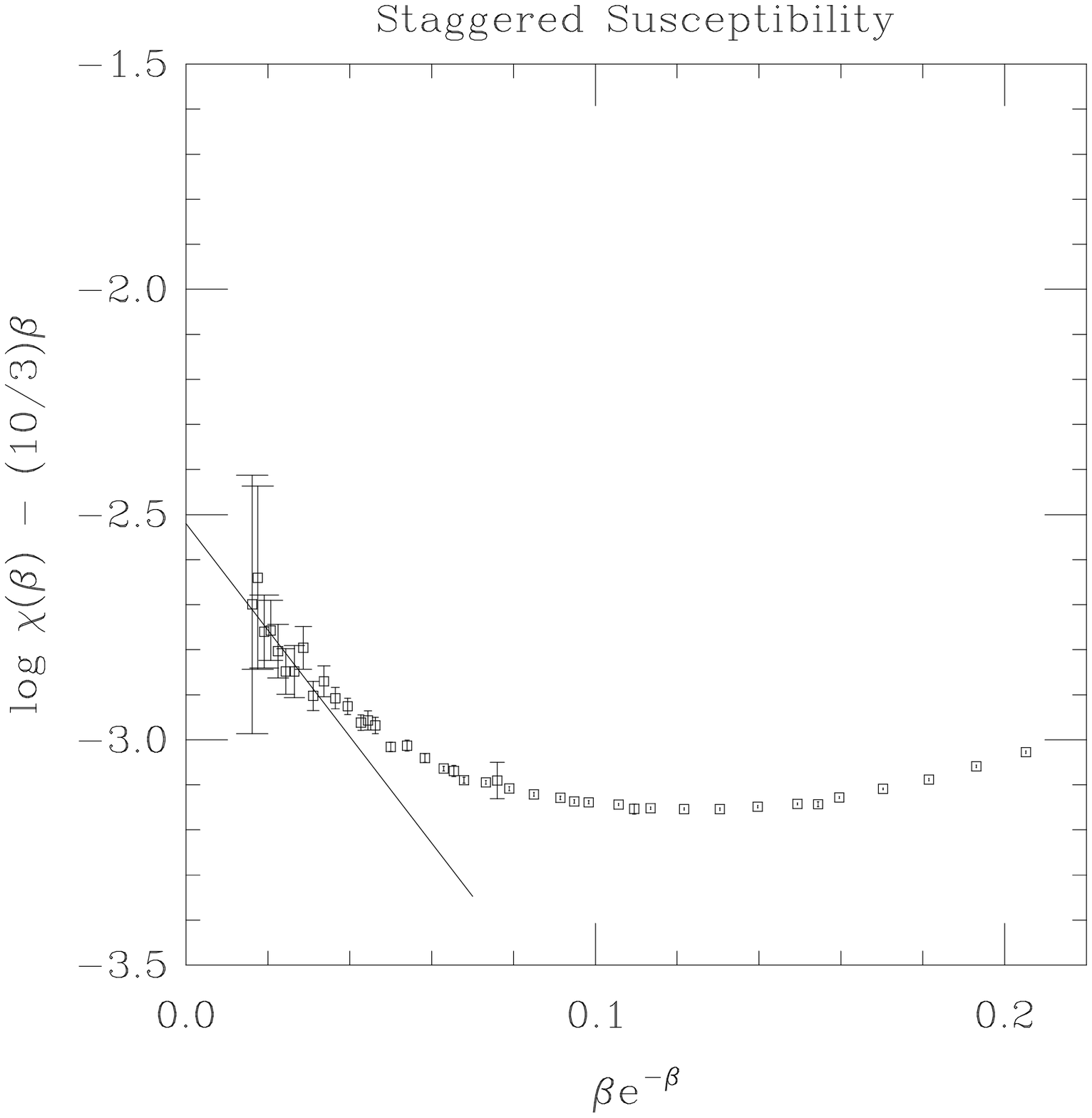}}
\vspace*{-1.5cm}
\caption{
   $\log \chi(\beta) - \ffrac{10}{3}\beta$ versus $\beta e^{-\beta}$.
   Straight line is
   $\log \chi(\beta) - \ffrac{10}{3}\beta = -2.52 - 11.82\beta e^{-\beta}$.
}
\label{fig.sec4.chiplot2}
\end{figure}

\begin{figure}[p]
\centerline{
\epsfxsize=4in
\epsfbox{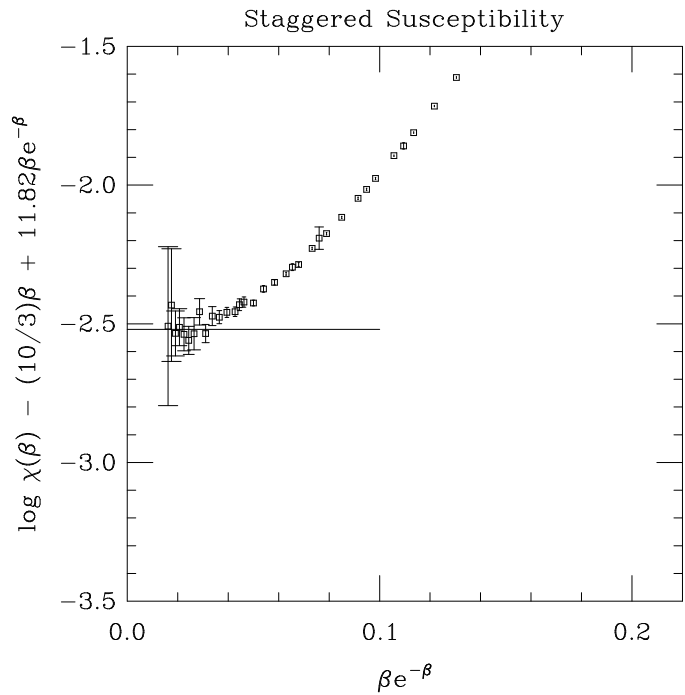}}
\vspace*{-1.5cm}
\caption{
   $\log \chi(\beta) - \ffrac{10}{3}\beta + 11.82\beta e^{-\beta}$
   versus $\beta e^{-\beta}$.
   Straight line is
   $\log \chi(\beta) - \ffrac{10}{3}\beta + 11.82\beta e^{-\beta} = -2.52$.
}
\label{fig.sec4.chiplot3}
\end{figure}

\begin{figure}[p]
\centerline{
\epsfxsize=4in
\epsfbox{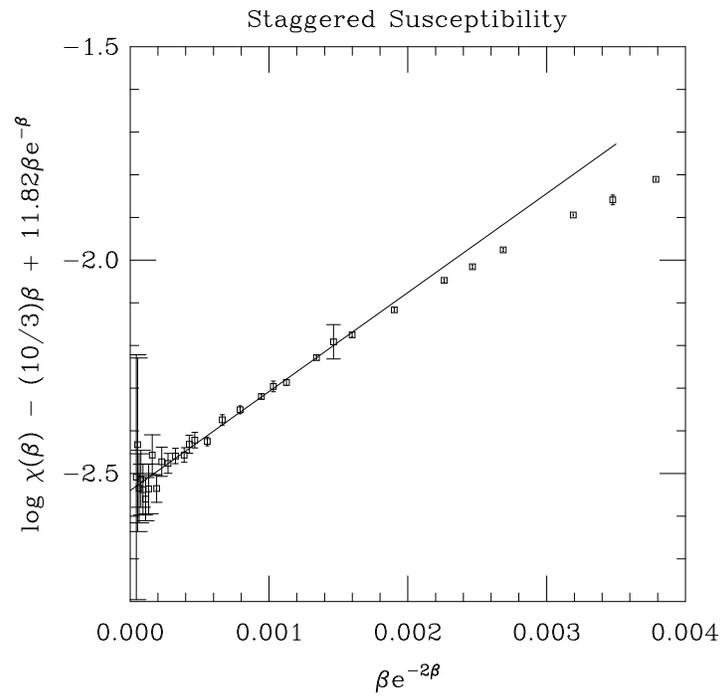}}
\vspace*{-1.5cm}
\caption{
   $\log \chi(\beta) - \ffrac{10}{3}\beta + 11.82\beta e^{-\beta}$
   versus $\beta e^{-2\beta}$.
   Straight line is
   $\log \chi(\beta) - \ffrac{10}{3}\beta + 11.82\beta e^{-\beta} =
    -2.54 + 232 \beta e^{-2\beta}$.
}
\label{fig.sec4.chiplot4}
\end{figure}

In the Coulomb gas picture of vortices interacting with a strength
$\propto K$, the tightly-bound vortex-antivortex pairs of vorticity $\pm3$ 
(Figure~\ref{fig1}a) act to screen this interaction,
thus decreasing the effective $K$.
A very similar effect happens in the Coulomb gas picture of the XY
model, and we show that a calculation similar to that employed in the
Kosterlitz renormalization-group approach \cite{Kosterlitz}
may be used here to estimate the shift in $K$,
to first order in $w=e^{-\beta}$. 

Rather than evaluating the interaction between two vortices, it is
simpler to consider the correction to the height correlation function,
which, in the absence of defects, has the form 
\begin{equation}
\label{hh}
\langle [h({\bf r}_1)-h({\bf r}_2)]^2\rangle \;\sim\;
   (\pi K)^{-1}\log|{\bf r}_1-{\bf r}_2|/a+{\rm const.}
\end{equation}
Introducing defects of the form 
\begin{equation}
h_{\rm def}({\bf r})=D{{\bf n}\cdot({\bf r}-{\bf r}_0)\over
                            |{\bf r}-{\bf r}_0|^2}
\end{equation}
where $\bf n$ is a unit lattice vector,
the first-order correction to the $\langle h({\bf r}_1) h({\bf r}_2) \rangle$
term in (\ref{hh}) is
\begin{equation}
\label{fo}
e^{-\beta}e^{-S_{\rm def}}D^2\sum_{{\bf r}_0,{\bf n}}
{({\bf n}\cdot({\bf r}_1-{\bf r}_0))({\bf n}\cdot({\bf r}_2-{\bf r}_0))
\over |{\bf r}_1-{\bf r}_0|^2 \, |{\bf r}_2-{\bf r}_0|^2}
\end{equation}
where $S_{\rm def}$ is the action (relative negative entropy) of the defect.
Summing over the four orientations of $\bf n$ turns the numerator into
$2({\bf r}_1-{\bf r}_0)\cdot({\bf r}_2-{\bf r}_0)$. 
In the continuum limit, (\ref{fo}) may be written
\begin{equation}
\label{fo2}
2 e^{-\beta}e^{-S_{\rm def}} D^2\int
{({\bf r}_1-{\bf r}_0)\cdot({\bf r}_2-{\bf r}_0)\over
|{\bf r}_1-{\bf r}_0|^2 \, |{\bf r}_2-{\bf r}_0|^2} \, d^2{\bf r}_0
\end{equation}
Apart from pieces which contribute to terms independent of
$|{\bf r}_1-{\bf r}_2|$,
the numerator in (\ref{fo2}) may be rewritten as 
$-\frac12[({\bf r}_1-{\bf r}_0)-({\bf r}_2-{\bf r}_0)]^2 =
 -\frac12({\bf r}_1-{\bf r}_2)^2$, so that
(\ref{fo2}) becomes
\begin{equation}
\label{fo3}
- e^{-\beta}e^{-S_{\rm def}} D^2\int{|{\bf r}_1-{\bf r}_2|^2\over
|{\bf r}_1-{\bf r}_0|^2 \, |{\bf r}_2-{\bf r}_0|^2} \, d^2{\bf r}_0
  \;\sim\; - e^{-\beta}e^{-S_{\rm def}}
    D^2\cdot 4\pi\log|{\bf r}_1-{\bf r}_2|/a
\end{equation}
(This last result may be most easily seen by evaluating the dependence
on the cut-off $a$ close to
${\bf r}_0\sim {\bf r}_1$ and ${\bf r}_0\sim {\bf r}_2$.)
{}From (\ref{fo3}) may be read off the first-order correction to the
stiffness constant
\begin{equation}
\delta K/K^2 \;=\; -2\pi\cdot D^2\cdot4\pi e^{-\beta}e^{-S_{\rm def}}   \;.
\end{equation}
Setting $K=\pi/6$ thus gives
\begin{equation}
\delta K=-(2\pi^4/9)\cdot D^2 \cdot e^{-\beta}e^{-S_{\rm def}}   \;.
\end{equation}

On the other hand, the first-order correction of such defects to the
free energy per vertex is
\begin{equation}
\delta f=-4e^{-\beta}e^{-S_{\rm def}}=-c_1e^{-\beta}
\end{equation}
where $c_1$ is the number appearing in (\ref{energy2})/(\ref{energy1}) and 
numerically determined to be $\approx 0.218$. This gives the
value $dK/dw|_{w=0}=-(\pi^4/18)D^2c_1$, leading to the estimate
$D\approx 0.70$. This is in rough
agreement with various rather crude estimates which may
be made on the assumption that the effective Hamiltonian $S_{\rm G}$
is valid down to the lattice scale.


\clearpage


\clearpage

\clearpage

\end{document}